\documentclass{article}

\usepackage{cite}
\usepackage{amsmath,amssymb,amsfonts}
\usepackage{algorithmic}
\usepackage{graphicx}
\usepackage{textcomp}
\usepackage{booktabs}
\usepackage{rotating}

\usepackage[a4paper, total={6in, 9in}]{geometry}

\begin{document}

\title{A Backside-Illuminated Charge-Focusing Silicon SPAD with Enhanced Near-Infrared Sensitivity}
\author{
	Edward~Van~Sieleghem$^{1,2}$, Gauri~Karve$^{1}$, Koen~De~Munck$^{1}$, Andrea~Vinci$^{1}$,\\
	Celso~Cavaco$^{1}$, Andreas~S\"uss$^{3}$, Chris~Van~Hoof$^{1,2}$, Jiwon~Lee$^{1}$\\
	\small$^1$Imec, 3001 Leuven, Belgium\\
	\small$^2$Department of Electrical Engineering, Katholieke Universiteit Leuven, 3001 Leuven, Belgium\\
	\small$^3$OmniVision Technologies, Santa Clara, CA 95054 USA\\
}
\date{}
\maketitle

\textbf{\textcopyright 2022 IEEE.  Personal use of this material is permitted.  Permission from IEEE must be obtained for all other uses, in any current or future media, including reprinting/republishing this material for advertising or promotional purposes, creating new collective works, for resale or redistribution to servers or lists, or reuse of any copyrighted component of this work in other works.\newline}


\begin{abstract}
A backside-illuminated (BSI) near-infrared enhanced silicon single-photon avalanche diode (SPAD) for time-of-flight (ToF) light detection and ranging applications is presented. The detector contains a 2~µm wide multiplication region with a spherically-uniform electric field peak enforced by field-line crowding. A charge-focusing electric field extends into a 10~µm deep absorption volume, whereby electrons generated in all corners of the device can move efficiently towards the multiplication region. The SPAD is integrated with a customized 130~nm CMOS technology and a dedicated BSI process. The device has a pitch of 15~µm, which has the potential to be scaled down without significant performance loss. Furthermore, the detector achieves a photon detection efficiency of 27\% at 905~nm, with an excess bias of 3.5~V that is controlled by integrated CMOS electronics, and a timing resolution of 240~ps. By virtue of these features, the device architecture is well-suited for large format ToF imaging arrays with integrated electronics.
\end{abstract}

\textbf{Keywords}: Single-photon avalanche diode (SPAD), backside illumination (BSI), CMOS integrated circuits, infrared imaging, charge focusing.\\

\textbf{Funding}: This work was supported in part by Research Foundation -- Flanders (FWO) SB PhD fellowship under Grant 11D3321N.\\

\textbf{Corresponding author}: Jiwon.lee@imec.be

\section{Introduction}
Single-photon avalanche diodes (SPADs) enable the detection of individual photons in the time domain by exploiting avalanche breakdown in an electric field. This time resolving capability finds its use in scientific and industrial applications, including biophotonics \cite{brushini2019}, quantum communications \cite{ceccarelli2021}, and time-of-flight (ToF) imaging \cite{morimoto2020}. SPADs are increasingly adopted for mobile and automotive light detection and ranging (LIDAR) systems based on ToF principles \cite{villa2021}. These systems typically operate at near-infrared (NIR) and short-wave infrared wavelengths such as 905~nm, 940~nm, and 1550~nm, for which solar background noise is reduced and lasers are readily available \cite{rablau2019}. LIDAR systems benefit from the integration of SPADs and electronic circuits in large arrays with high NIR photon detection efficiency (PDE) \cite{zhang2018}.

Silicon SPADs for the detection of NIR radiation have been researched extensively \cite{ceccarelli2021}. The benefits of silicon detectors include: (i) fabrication in well-established manufacturing platforms, (ii) ease of integration with CMOS circuits, and (iii) low detector noise at room temperature \cite{charbon2014}. In particular, the low crystal defectivity and relatively high band gap of silicon result in reduced dark count rate (DCR) contributions from trap-assisted generation and tunneling. However, the larger band gap also leads to a relatively low absorption coefficient for NIR photons \cite{sze1969}. Due to the inefficient absorption of infrared light, silicon SPADs may require an absorption volume with a depth exceeding 5~µm to achieve a PDE above 10\% at 905nm \cite{ito2020,vansieleghem2021,gulinatti2021,jegannathan2020,acerbi2018silicon,takai2016single,stipcevic2013,webster2012high}.

A typical NIR-sensitive silicon SPAD contains a planar p-n junction embedded in a thick lowly-doped or graded absorption volume \cite{ceccarelli2021}. When the device operates at a finite excess bias above the breakdown voltage, the p-n junction contains a laterally uniform electric field peak. This field peak constitutes the multiplication region of the device. Charge carriers generated inside and below the multiplication region can trigger discrete avalanche breakdown events. The breakdown triggering probability relates to the strength of the multiplication field, which increases gradually with the excess bias \cite{zappa2007}. A guard ring and other doped regions are located near the edges of the planar junction to prevent premature breakdown, establish electrical contacts, and provide isolation. Carriers generated in this peripheral volume have a low probability of moving through the field peak and triggering breakdown \cite{gulinatti2021,acerbi2018silicon}. The corresponding loss in the detection efficiency is more substantial for small SPADs, in which the peripheral region is proportionally larger. Therefore, to scale down silicon SPADs without significant impact on the PDE \cite{morimoto2020gardring}, further device improvements are required.

A frequently used technique for boosting the PDE is backside illumination (BSI) \cite{lindner2017high,abbas2016backside}. Additionally, optical microlenses can be introduced for focusing light into the volume below the multiplication region \cite{ito2020}. Furthermore, SPADs have been demonstrated containing charge-focusing electric fields that funnel charges generated in the peripheral volume, including the guard ring, into the multiplication region \cite{vansieleghem2021,jegannathan2020,iwata2020lensing}. In these devices, carriers generated close to the edges of the SPAD can also reach the multiplication field, which significantly enhances the PDE. A near-unity fill factor can be achieved irrespective of the detector pitch. Charge focusing functionality has also recently been demonstrated in the domain of silicon photomultipliers (SiPMs) \cite{engelmann2020tip}.

Another consideration for NIR silicon SPADs is the photon timing resolution. The deep absorption region potentially results in significant variability of the transport time of photogenerated carriers. To attain an acceptable timing resolution, an electric field must be present in the absorption volume, for example, by depletion. Thick NIR-sensitive SPADs with depleted absorption volumes generally achieve a photon timing resolution between 100~ps and 300~ps, which is higher than thin SPADs \cite{zappa2007}. If the depletion region becomes too deep, the electric field strength may become insensitive to the reverse bias; When the applied potential has to drop uniformly over a large distance, the field strength does not increase rapidly with the reverse voltage. Consequently, the sensor may require a high excess bias, above 10~V, to achieve an optimal breakdown triggering probability \cite{gulinatti2021}. Such excess bias is more difficult to control with integrated CMOS electronics \cite{lindner2017high}. Instead, less compact customized electronics are sometimes needed \cite{acconcia2016}. In general, careful design optimization is required to achieve a good temporal response, a high NIR PDE, a small form factor, and CMOS compatibility \cite{ito2020}.

In this work, we present a BSI silicon SPAD with state-of-the-art NIR efficiency. The device architecture is based on a previously reported frontside illuminated (FSI) SPAD \cite{vansieleghem2021}. The detector contains a deep depleted absorption volume with a charge-focusing electric field. Most photo-generated carriers are funneled towards a small multiplication region where they can trigger breakdown. Similar to other charge-focusing SPADs, the device can be easily scaled and integrated into arrays \cite{iwata2020lensing}. Unlike those SPADs, the multiplication field is enforced by field-line crowding instead of a steep p-n junction, and the field strength decreases gradually into the absorption volume. The field-line crowding effect is achieved with a conventional planar processing technology. A competitive timing resolution is attained, and the detector operates efficiently at a low excess bias controlled by CMOS electronics. These features are promising for large format ToF imager arrays.

\section{SPAD device}
\label{sec:spad_device}
\begin{figure}
	\centerline{\includegraphics[width=4.0in]{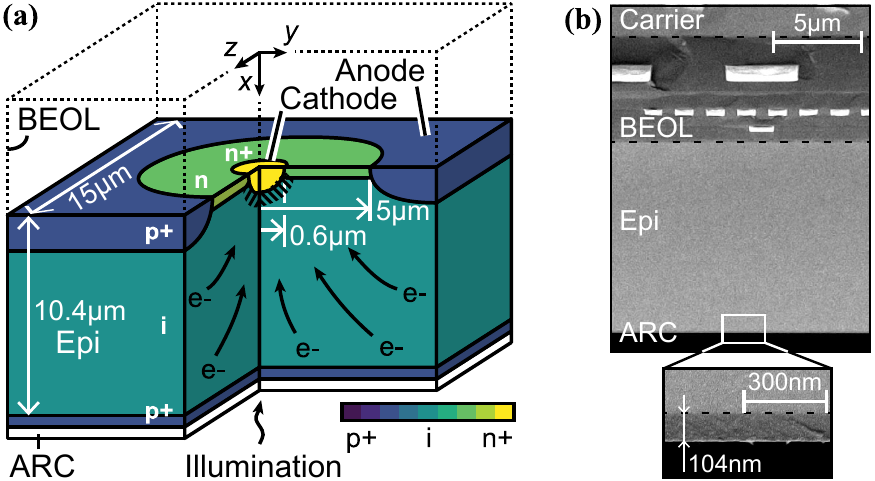}}
	\caption{(a) Doping profile of a BSI SPAD. Electrons (e-) drift to a central field peak (dashed area). (b) X-SEM images of the wafer.}
	\label{fig:device}
\end{figure}

The BSI SPAD was designed based on the architecture of a previously reported FSI device \cite{vansieleghem2021}, and it was manufactured in a customized 130~nm CMOS technology. The backside processing was executed roughly according to the flow described by B.~Vereecke \textit{et al.} \cite{vereecke2015}. Fig.~\ref{fig:device}(a) illustrates the structure and doping profile of the SPAD. The detector has a size of 15~µm and contains a p-intrinsic volume (Epi) with a depth of 10.4~µm and a doping concentration of $\approx10^{12}$~cm$^{-3}$. The cathode is formed by a central n-type region with a hemispherical surface and a radius of $\approx0.6$~µm. The anode consists of a p-type region that encloses the cathode with an inner radius of 5~µm. The top oxide-silicon interface between the cathode and anode contains a shallow n-type implantation. The backside is passivated by a thin p-type doped region and contains a silicon-nitride anti-reflective coating (ARC) of approximately 100~nm thickness. The backside doped layer is electrically connected to the anode potential.

Fig.~\ref{fig:device}(b) shows a cross-section scanning electron microscope (X-SEM) image of the layers in the device. The backend of line (BEOL) is located between the silicon substrate and a carrier wafer (Carrier), and it contains the metal interconnects (bright patches). To enhance the PDE, a metal reflector that covers approximately half of the detector area is incorporated in the BEOL. This reflector is not visible in the figure because the X-SEM was taken outside of the pixel area.

\begin{figure}
	\centerline{\includegraphics[width=4.0in]{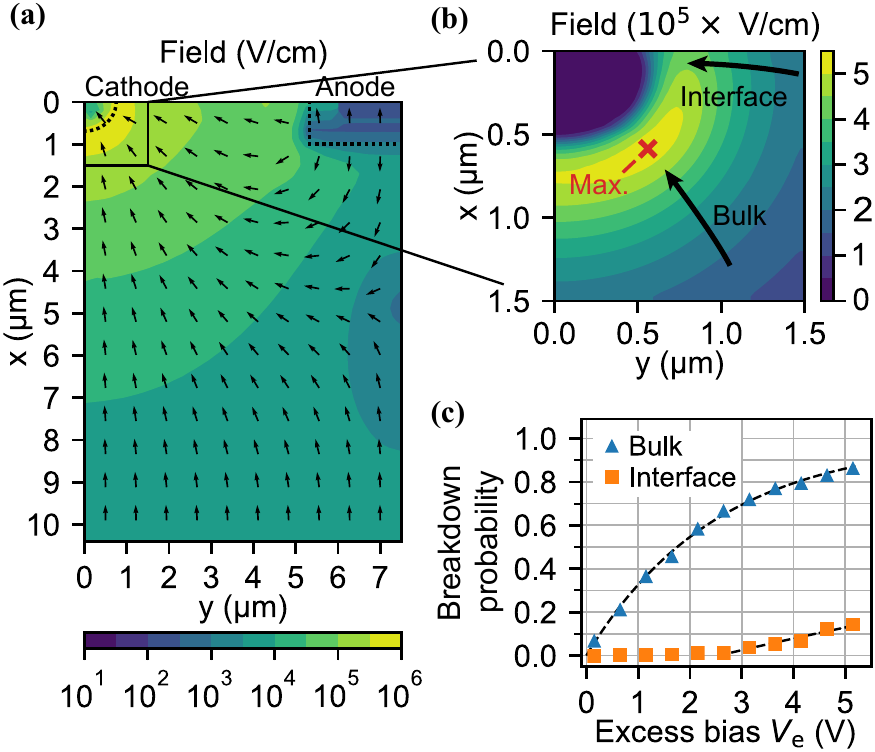}}
	\caption{(a) Simulated electric field and electron flow in a cross-section of the device at $V_\mathrm{e}=3.5$~V. (b) Zoom of the field peak. (c) Simulated breakdown probability for electrons entering the multiplication region.}
	\label{fig:field}
\end{figure}

Fig.~\ref{fig:field}(a) demonstrates the simulated electric field in a cross-section of the device when the SPAD is operated at an excess bias $V_\mathrm{e}=3.5$~V beyond the breakdown voltage $V_\mathrm{bd}$. The figure includes the flow vectors for electrons. The intrinsic volume and shallow n-implantation are depleted. The space charge region (SCR) extends to the backside and contains a charge-focusing electric field. Electrons generated in the entire device volume of 15~µm~$\times$~15~µm~$\times$~10~µm move quickly toward the cathode by drift, as is evidenced by the flow vectors. Furthermore, the density of the field lines is higher near the cathode, resulting in a multiplication region with a diameter of $\approx2$~µm. The multiplication region contains a spherically-uniform electric field peak, as demonstrated in Fig.~\ref{fig:field}(b). Electrons entering the multiplication region from most directions experience a uniform probability of triggering breakdown. Hence, the charge-focusing phenomenon is very efficiently implemented. Besides, the field peak is locally reduced on the top interface by the shallow n-implantation. Carriers moving horizontally along the top interface experience a much lower breakdown probability due to the reduced field strength.

Fig.~\ref{fig:field}(c) shows the simulated breakdown triggering probability for electrons arriving at the multiplication region from the bulk, moving through the indicated field maximum, and for electrons moving along the top interface. The figure was created by simulating the transport and multiplication of carriers with a Monte Carlo tool \cite{vansieleghem2021iisw}. The breakdown triggering probability for electrons generated in the bulk increases rapidly with the excess bias, with a value $>70$\% at 3.5~V. In contrast, the breakdown probability on the interface remains relatively low. This behavior is beneficial for the PDE and the DCR: Photo-generated carriers are likely to be detected since most of them originate from the bulk. Thermally generated carriers generated on, or near, the defective interface are less likely to cause breakdown events because they tend to travel parallel to the interface.

Unlike typical SPADs, the multiplication field is not formed by ionized doping atoms in a p-n junction. Instead, this region is created because of field-line crowding in the intrinsic volume around the cathode; Due to the geometry of the device, the density of the electric field lines is highest in this region. Although most of the applied potential drops close to the cathode, the field also extends into the absorption volume. An expression for the electric field strength $E$ can be derived from Gauss's law. The derivation assumes a highly doped, spherical cathode that transitions abruptly into the intrinsic layer. The cathode is also assumed to be much smaller than the other device dimensions. Under these conditions, the field at distance $r$ from the cathode center is approximately
\begin{equation}
	\label{eq:field}
	E \approx
	\begin{cases}
		\dfrac{(V_\mathrm{r} + V_\mathrm{bi})r_\mathrm{n}}{r^2} & \quad \text{if} \quad r_\mathrm{n} < r \\
		0 & \quad \text{otherwise}
	\end{cases},
\end{equation}
with reverse bias $V_\mathrm{r}$, built-in voltage $V_\mathrm{bi}$, and cathode radius $r_\mathrm{n}$ \cite{jackson1999}. The field is peaked at radius $r_\mathrm{n}$ and decays at a rate of $r^{-2}$. Unlike other works that exploit field-line crowding \cite{engelmann2020tip}, the device is fabricated in a planar technology.

Following equation~\ref{eq:field}, a smaller cathode radius results in a sharper field peak. Therefore, the breakdown- and operating voltages can be reduced by decreasing the size of the cathode. At the lower operating voltage, the electric field in the absorption volume is also weaker. Consequently, a reduction of the breakdown voltage may lead to a degradation of the timing resolution. The timing loss can be recovered by reducing the detector pitch and depth, such that carriers generated in device corners have shorter travel distances.

In this work, a breakdown voltage of $\approx67$~V is achieved. This value is relatively high, which can negatively affect power consumption and ease of integration \cite{cova1996}. The breakdown voltage of the presented device embodiment was not reduced further due to practical constraints in the fabrication process. The doping conditions and cathode shape could not yet be optimized to achieve a lower breakdown voltage.

Based on equation \ref{eq:field} and numerical simulations, the authors estimate that a breakdown voltage less than 30~V can be attained by decreasing the cathode radius. If the pitch and depth of the device are also scaled down to values between 5~µm and 7~µm, a timing resolution well below 150~ps can likely be obtained while maintaining a high NIR sensitivity. At an operating voltage below 30~V, the field peak near the cathode can become very sharp, and special care must be taken to prevent additional band-to-band tunneling noise.

\section{Array and circuit integration}
\begin{figure}
	\centerline{\includegraphics[width=4.0in]{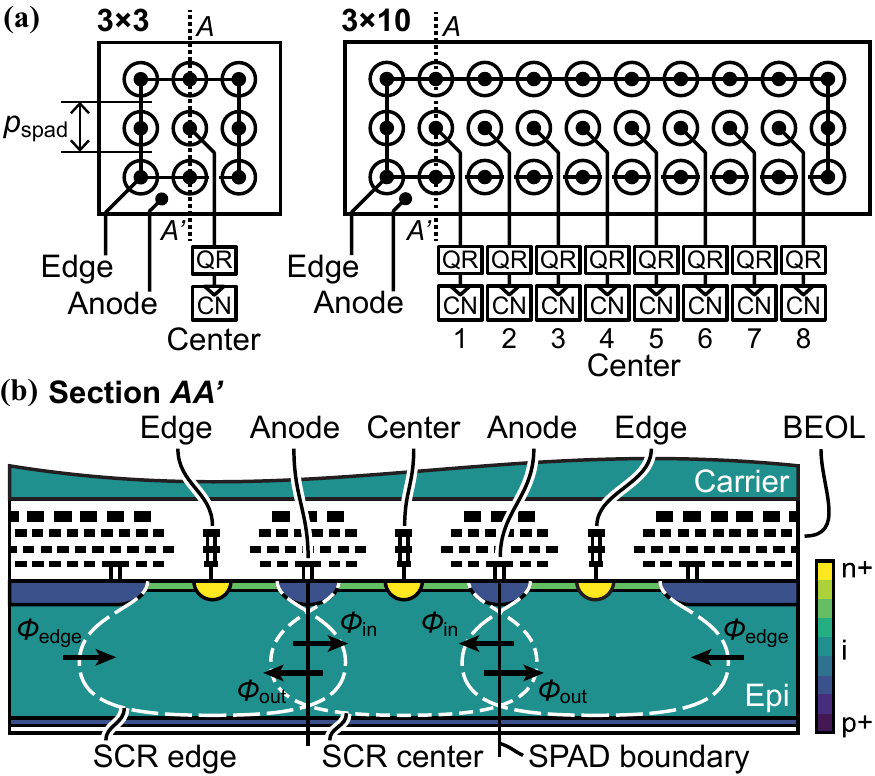}}
	\caption{(a) Schematics of SPAD arrays on which electrical connections and circuits are indicated. (b) Doping profile cross-section in which SCRs (dashed lines) and charge fluxes (arrows) are indicated.}
	\label{fig:section}
\end{figure}

The SPAD is integrated into uniform arrays with pitch $p_\mathrm{spad}=15$~µm, as demonstrated in Fig.~\ref{fig:section}(a). On the one hand, the $3 \times 3$ arrays contain one central SPAD surrounded by a ring of eight edge diodes. On the other hand, the $3 \times 10$ arrays incorporate a line of eight central SPADs surrounded by twenty-two edge diodes. All devices in the arrays share a common anode potential. Likewise, the cathodes of the edge diodes are connected to a common voltage source. The cathodes of the central SPADs are connected to independent monolithic integrated circuits for active quenching and recharging (QR). The circuits also contain digital counters (CN) and output buffers for monitoring breakdown events. Reference \cite{vansieleghem2021} describes the operation of the readout. Besides, for the purpose of current-voltage (IV) characterization, array modules are also available in which the central cathodes are directly connected to contact pads without integrated readout.

The SPAD performance is evaluated exclusively by measuring the breakdown events of the central SPADs in arrays. The purpose of the edge diodes is to control the electrostatic environment of the central SPADs such that measurements thereof are not affected by peripheral effects. The central devices are operated at an excess bias $V_\mathrm{e}$ above the breakdown voltage, and the edge diodes are biased at a voltage between 0.5~V and 1.5~V below $V_\mathrm{bd}$. Following equation \ref{eq:field}, most of the applied potential drops close to the cathode regions, and the field strength decreases rapidly with distance in the epi. Therefore, the SCRs in the center- and edge diodes are nearly identical, despite the different reverse biases. Consequently, the central SPADs more or less experience the environment of an active SPAD array and can be characterized accordingly. Furthermore, measurements are not tainted by breakdown in the edge diodes. There is also a large potential barrier between neighboring cathodes, whereby punch-through is prevented.

To further demonstrate the effect of the edge diodes, Fig.~\ref{fig:section}(b) illustrates a cross-section of an array. Each dashed line in the figure indicates the outer SCR-edge of a diode, assuming that it is the only device that is being biased. When all diodes are operated simultaneously, the SCRs merge at the vertical boundaries between neighboring devices. Some charges that are generated near the boundaries can diffuse from one diode to another, resulting in electrical crosstalk. The corresponding carrier fluxes in and out of the central SPAD are indicated with vectors $\varPhi_\mathrm{in}$ and $\varPhi_\mathrm{out}$, respectively. Since all diodes contain nearly-identical weak electric fields in the absorption volume, and the generation is approximately the same for all devices, the in- and out-fluxes are nearly identical in magnitude. As a result, the net flux is approximately zero, and the characterization results of the central SPADs are not biased by electrical crosstalk. Furthermore, the net charge flux $\varPhi_\mathrm{edge}$ from the periphery of the array also does not bias the measurements because this flux is largely collected by the edge diodes. These claims have been verified experimentally by further decreasing the edge-diode bias beyond the nominal operating voltage. No significant change was observed in the characterization results up to a voltage difference $\gg 5$~V.

Unlike the $3 \times 3$ arrays, the detectors in $3 \times 10$ arrays can also experience breakdown-induced optical crosstalk. Photons generated in the charge cloud of a primary breakdown event can be absorbed in a nearby SPAD, where they can trigger secondary breakdown events \cite{zappa2007}. The crosstalk probability $\chi_n$ is the chance that optical crosstalk occurs between SPADs at distance $n \times p_\mathrm{spad}$ during breakdown events. Because the SPAD arrays of this work contain deep absorption volumes and connected depletion regions, the crosstalk probability is expected to be relatively high \cite{ito2020}.

Finally, a form of spatial optical crosstalk (SOC) can exist. Unlike breakdown-induced optical crosstalk, SOC involves the incident photon flux; Optical rays that enter one SPAD can cross the vertical boundaries into another SPAD. This form of crosstalk is worse when the absorption volume is deeper and when the illumination is more angled. An interplay can also exist between electrical crosstalk and SOC, where photo-carriers generated near SPAD boundaries diffuse into neighboring devices due to the weak field. The quantitative analysis of this phenomenon is beyond the scope of this work. Optical microlenses and deep trench isolation (DTI) can be employed to compensate for the different forms of crosstalk \cite{agranov2003,ito2020}.

\section{Results}
In this section, the SPAD performance is reported. Unless specified otherwise, the performance was extracted from the central devices of $3 \times 3$ arrays. All measured performance metrics were reproducible and uniform across a wafer.

\subsection{Quasi-static behavior}
\begin{figure}
	\centerline{\includegraphics[width=4.0in]{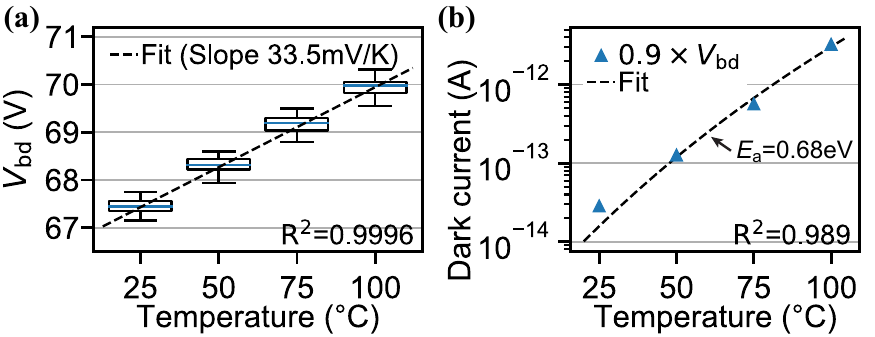}}
	\caption{Breakdown voltage (a) and median dark current (b) extracted from the IV characteristics of 29 devices on a wafer.}
	\label{fig:iv}
\end{figure}

The quasi-static IV characteristics of 29 devices across a wafer were measured. The breakdown voltage, series resistance, and dark current were extracted. Fig.~\ref{fig:iv}(a) shows the distribution of $V_\mathrm{bd}$ for different temperatures. The median breakdown voltage is 67.4~V at 25~°C, with a temperature coefficient of 33.5~mV/K and a minimum-maximum range of 0.6~V. The temperature coefficient is comparable to previously reported avalanche diodes \cite{acerbi2018silicon}. A median series resistance of 53~kOhm is extracted from the slopes of the IV curves in the breakdown condition. The main contributors are likely the resistances of the space-charge region and the neutral regions \cite{charbon2014}. The high resistance value can have a negative effect on power consumption \cite{cova1996}. However, no evidence was found that it degrades the quenching behavior or efficiency of the device. The RC delay constant for quenching remains acceptable due to the low junction capacitance of 0.4~fF (simulated).

Fig.~\ref{fig:iv}(b) illustrates the dark current versus temperature at a reverse bias of $0.9 \times V_\mathrm{bd}$. The current equals $\approx30$~fA at 25~°C, which is close to the detector noise floor. The fitted activation energy is 0.68~eV, which is consistent with Shockley-Read-Hall (SRH) generation, assisted by defects \cite{zappa2007}. The primary location of the defects is likely the top oxide-silicon interface. The defect density introduced during epitaxial growth and doping implantations is expected to be relatively low.

\subsection{Dark count rate}
\begin{figure}
	\centerline{\includegraphics[width=4.0in]{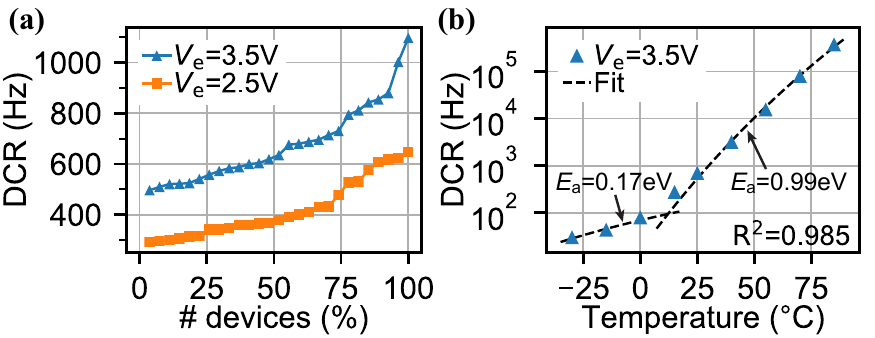}}
	\caption{(a) Cumulative DCR distribution for 27 devices at room temperature. (b) DCR versus temperature of a selected device at $V_\mathrm{e}=3.5~V$.}
	\label{fig:dcr}
\end{figure}

The dark count rates of 27 devices across a wafer were measured. Fig.~\ref{fig:dcr}(a) demonstrates the cumulative DCR distribution at room temperature, with median values of 380~Hz and 640~Hz at excess biases 2.5~V and 3.5~V, respectively. Fig.\ref{fig:dcr} (b) presents the temperature dependence of the DCR for a selected device at $V_\mathrm{e}=3.5$~V. The data are fitted by the sum of two Arrhenius equations. At temperatures above 15~°C, the DCR activation energy is 0.99~eV, which is consistent with thermal generation. This value is 0.31~eV higher than the activation energy of the dark current, suggesting that SRH-generated carriers have a low probability of multiplying; Electrons that are generated on the defective top interface are not likely to move through the field maximum and trigger breakdown. At temperatures below 15~°C, the DCR activation energy is 0.17~eV, which is consistent with temperature-insensitive tunneling \cite{webster2012high}. The electric field close to the cathode is sharp, likely resulting in band-to-band tunneling.

\subsection{Afterpulsing}
\begin{figure}[!t]
	\centerline{\includegraphics[width=4.0in]{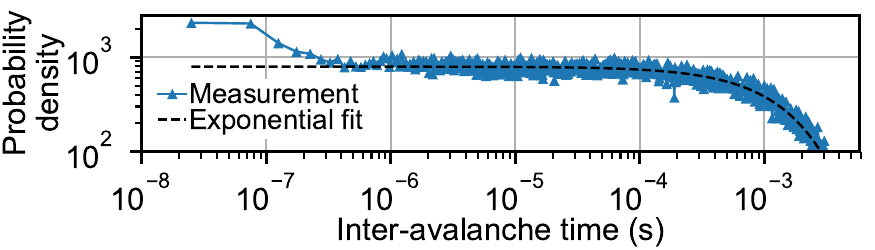}}
	\caption{Inter-avalanche time histogram and exponential fit with $V_\mathrm{e}=3.5$~V, 50~ns bin width, and sample size $2.5 \times 10^6$.}
	\label{fig:ap}
\end{figure}

To evaluate the afterpulsing probability, the time between consecutive avalanche events was measured for selected devices. Fig.~\ref{fig:ap} shows the inter-avalanche time histogram of a SPAD with $V_\mathrm{e}=3.5$~V, a temperature of 25~°C, and a dead time of 14~ns. The curve is fitted by an exponential function. The area between the measurement and the fit corresponds to afterpulsing \cite{charbon2014} with a probability of $<0.1$\%. This low value is likely achieved because of the near absence of defects in the small multiplication region.

\subsection{Photon detection efficiency}
\begin{figure}
	\centerline{\includegraphics[width=4.0in]{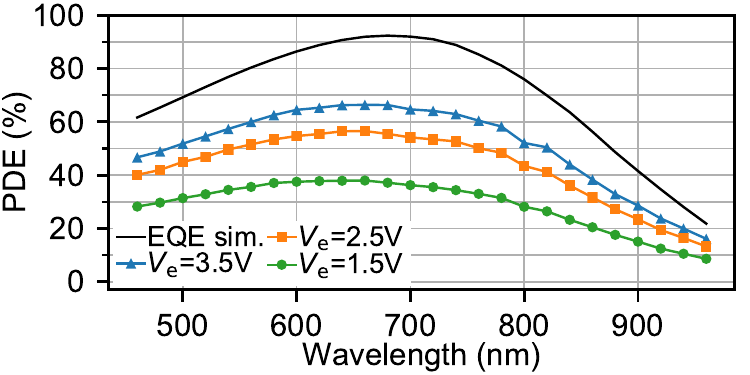}}
	\caption{PDE for different excess biases and simulated EQE.}
	\label{fig:pde}
\end{figure}

The photon detection efficiency was measured by illuminating a SPAD with a setup consisting of a xenon lamp, a monochromator, and an integrating sphere. Fig.~\ref{fig:pde} presents the measured PDE as a function of wavelength $\lambda$ at different excess biases. The PDE has a peak of 66\% near 660~nm, and a value of 27\% at 905~nm, for $V_\mathrm{e}=3.5$~V. The PDE is enhanced by the charge-focusing phenomenon, the deep absorption volume, and reflections in the BEOL. The relative measurement error is expected to be lower than 5\%.

Fig.~\ref{fig:pde} also shows the simulated external quantum efficiency (EQE) of the device. To obtain this result, the optical generation was simulated in a 3D representation of the device, including the ARC, top oxide interface, and BEOL metals. The PDE is equal to the product of the EQE and the average breakdown probability \cite{charbon2014}. By dividing the measured PDE at $V_\mathrm{e}=3.5$~V by the simulated EQE, it becomes evident that an average breakdown probability between 70\% and 75\% is achieved over a broad range of wavelengths. This probability is in line with the simulation results of Fig.~\ref{fig:field}(c), considering most photo-generated electrons in the bulk have uniform breakdown probabilities.

\subsection{Temporal response}
\begin{figure}
	\centerline{\includegraphics[width=4.0in]{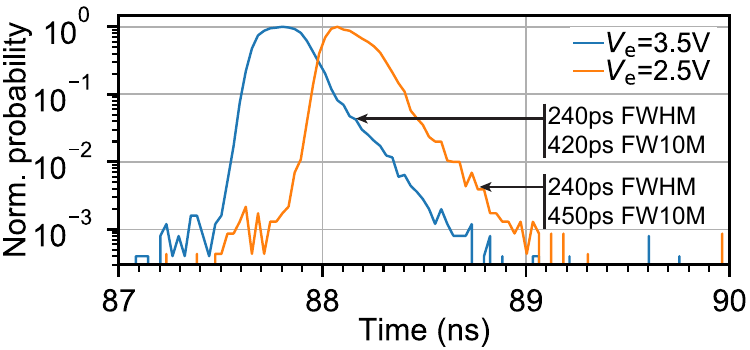}}
	\caption{Temporal response at different excess biases for $\lambda=905$~nm.}
	\label{fig:timing}
\end{figure}

The temporal response of the SPAD determines the resolution at which incident photons can be detected in the time domain. The response was measured by illuminating the device with an AUREA Technology Pixea picosecond pulsed laser at wavelength 905~nm; attenuated to $<0.1$ photons per pulse and with a repetition rate of 10~MHz. The time differences between the laser trigger signals and SPAD breakdown events were measured and binned with a resolution of 30~ps. Fig.~\ref{fig:timing} shows the measured temporal response for the SPAD at different excess biases. The result for $V_\mathrm{e}=3.5$~V has a full width at half maximum (FWHM) of 240~ps, a full width at tenth maximum (FW10M) of 420~ps, and no significant diffusion tail. The response for $V_\mathrm{e}=2.5$~V is similar. For both reported excess bias conditions, the timing resolution is expected to be limited by the transport of electrons in the absorption volume. At lower excess biases, the timing uncertainty is likely increased due to variance in the carrier multiplication process.

\subsection{Optical crosstalk}
\label{sec:crosstalk}
\begin{figure}
	\centerline{\includegraphics[width=4.0in]{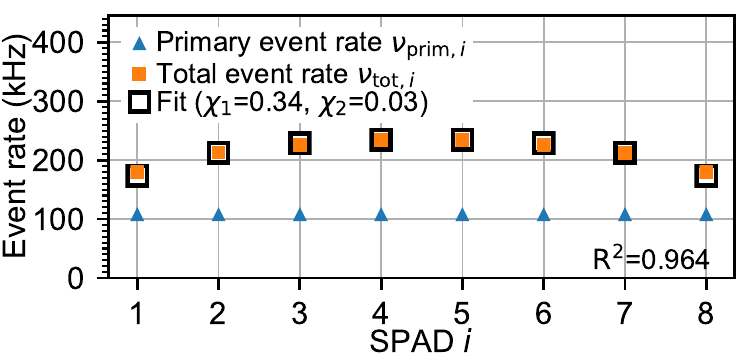}}
	\caption{Event rates in the central SPADs of $3~\times~10$ illuminated arrays, fitted by the crosstalk model of Appendix \ref{ap:crosstalk}.}
	\label{fig:crosstalk}
\end{figure}

The breakdown-induced optical crosstalk was analyzed with $3~\times~10$ arrays. Two different event rates were measured for the eight central SPADs in each array: The total event rate $\nu_{\mathrm{tot},i}$ and the primary event rate $\nu_{\mathrm{prim},i}$, where $i$ is the incremental index of the SPAD. The total event rate was measured while enabling all SPADs in each array simultaneously, whereas the primary event rate was extracted while activating only one SPAD at a time. Fig.~\ref{fig:crosstalk} shows the mean event rates based on nine homogeneously-illuminated test modules. The primary rates are identical for each SPAD ($\approx100$~kHz) due to the illumination. The total event rates decrease from the center outwards; Outer SPADs have fewer nearest neighbors and lower crosstalk. The crosstalk probabilities $\chi_1=0.34$ and $\chi_2=0.03$ are extracted by fitting the analytical model for the event rates described in appendix \ref{ap:crosstalk}. The obtained probabilities are relatively high due to the large depleted absorption volume and the absence of isolation between SPADs.

\section{Discussion}

\begin{sidewaystable*}
	\centering
	\caption{Performance of state-of-the-art silicon SPADs integrated with planar technologies}
	\begin{tabular}{|l|cccc|cc|}
		\hline
		& \multicolumn{4}{|c|}{Recent charge-focusing SPADs} & \multicolumn{2}{c|}{Recent NIR SPADs}\\
		& \textbf{This work} & \textbf{Van Sieleghem} & \textbf{Jegannathan} & \textbf{Iwata} & \textbf{Gulinatti} & \textbf{Ito} \\
		& & 2021 \cite{vansieleghem2021} & 2020 \cite{jegannathan2020} & 2020 \cite{iwata2020lensing} & 2021 \cite{gulinatti2021} & 2020 \cite{ito2020} \\
		\hline
		Technology				& BSI 130 nm	& FSI 130 nm 	& FSI 350 nm	& FSI 180 nm	& FSI Custom	& BSI 90 nm \\
		SPAD size				& 15 µm			& 15 µm			& 30 µm			& 14 µm			& 50 µm			& 10 µm \\
		Epi thickness			& 10.4~µm		& 18~µm			& 14 µm			& -				& 10 µm			& 7 µm \\
		$V_\mathrm{bd}$+$V_\mathrm{e}$ & 67.4+3.5 V & 67.3+3.5 V & 48.3+2.5 V	& 26.5+2.5 V	& 30+20 V		& 20+3 V \\
		PDE @ 905~nm			& 27\%			& 13\%			& $<$11\%		& 6\%			& 20\%			& 20\% \\
		DCR	@ RT				& 640 Hz		& 840 Hz		& $4\times10^6$ Hz & 3 Hz		& 3300 Hz		& 3 Hz \\
		Timing FWHM  @ $\lambda$ & 240 ps @ 905 nm & 350 ps @ 905 nm & 229 ps @ 840 nm & -		& 95 ps	@ 820 nm & 173 ps @ - \\
		Afterpusling			& $\leq$0.1\%	& $\leq$0.1\%	& 13\%			& -				& 2\%			& 0.1\% \\
		Crosstalk				& 34\%			& -				& -				& -				& 0.2\%†		& 0.3\%‡ \\
		\hline
		\multicolumn{7}{c}{†For SPAD pitch $\geq$250~µm, ‡20\% crosstalk probability when physical isolation omitted.} \\
	\end{tabular}
	\label{tab:comparison}
\end{sidewaystable*}

Table \ref{tab:comparison} outlines the performance of the presented BSI device and other silicon SPADs that employ charge focusing \cite{vansieleghem2021,jegannathan2020,iwata2020lensing} or that have a high NIR sensitivity \cite{gulinatti2021,ito2020}. Only SPADs are considered that are fabricated with a planar technology and that are intended for active array integration (unlike some SiPMs \cite{engelmann2020tip}). The present work compares favorably to the other publications. In particular, the measured PDE of 27\% at 905~nm is higher compared to those works. The high efficiency is achieved with an absorption depth of 10.4~µm, which is larger than some of the presented alternatives. Furthermore, the reported timing resolution is comparable to other devices and well-suited for LIDAR applications \cite{villa2021}.

Similar to other charge-focusing SPADs, carriers generated in a large absorption volume are efficiently funneled into a small multiplication region \cite{iwata2020lensing}. The advantages of this feature are as follows (in no particular order). (i) Firstly, the device dimensions can be scaled with relative ease. The scaling is limited by the requirement for the depletion of the entire epi layer. Since the field expands radially outwards from the cathode, full depletion can only be ensured when the pitch and depth of the detector have the same order of magnitude. (ii) Secondly, compared to SPADs with wider multiplication regions \cite{gulinatti2021,ito2020}, the device exhibits a lower afterpulsing probability. There are fewer defects in the multiplication region that facilitate afterpulsing due to the smaller area thereof \cite{jegannathan2020}. (iii) Thirdly, unlike typical SPADs with planar p-n junctions, optical microlenses are not required to focus carriers into the small multiplication region \cite{ito2020}. Nevertheless, such lenses can still be beneficial for improving performance, including crosstalk. (iv) Finally, the multiplication region is small. As a result, the device exhibits an exceptionally low junction capacitance of 0.4~fF. The low capacitance helps to speed up the quenching process, whereby the number of charges per breakdown event can be reduced \cite{cova1996}. Consequently, parameters such as the afterpulsing, power consumption, and timing response are potentially improved.

The presented SPAD also exhibits some features unique to the field-line crowding phenomenon that governs the field formation. (i) Firstly, the field is peaked near the cathode and gradually decays into the absorption volume. No dedicated p-n junction is required for the formation of the multiplication region. Unlike other SPADs with deep, depleted absorption volumes \cite{gulinatti2021}, the multiplication field remains sensitive to the reverse bias, and the device can operate efficiently at an excess bias below 3.5~V. (ii) At the same time, the field also remains sufficiently strong throughout the deep absorption volume to achieve a competitive timing resolution. Compared to typical NIR-sensitive SPADs, the trade-off between low excess bias and high drift field appears to be less pronounced \cite{gulinatti2021,ito2020}. Based on numerical simulations, the authors hypothesize that a scaled version of the device, with dimensions similar to the SPAD of Ito \textit{et al.} \cite{ito2020}, can achieve a timing resolution well below 150~ps and a PDE above 20\% at 905~nm. Such a device was not yet achieved in this work due to practical limitations and lack of optimization. (iii) Finally, the SPAD achieves good performance uniformity across dies and wafers. There are only a few process steps that can introduce non-uniformities.

It should be noted that a SiPM was recently presented by Engelmann \textit{et al.} that also employs the field-line crowding effect to achieve similar advantages \cite{engelmann2020tip}. Unlike this device, the SPAD of this work is fabricated with a simple planar processing technology, is intended for active SPAD array integration, can be easily operated with CMOS electronics, and is backside illuminated.

The presented device performs significantly better than the FSI implementation from which it is derived \cite{vansieleghem2021}. (i) Firstly, the timing resolution is higher, and the temporal response does not contain a diffusion tail due to the absence of large neutral regions. (ii) Secondly, the PDE is more than doubled because incident photons do not reflect from the BEOL metals before reaching the absorption volume. Instead, the metals in the BEOL now enhance the effective photon absorption depth. (iii) Finally, the DCR is slightly reduced due to the suppression of a previously unidentified leakage current from the periphery of the SPAD. Moreover, the backside processing does not appear to negatively affect the DCR.

Some parameters of the BSI SPAD are less optimal compared to previous publications. (i) Firstly, the breakdown voltage is high. This is not a fundamental limitation of the device. As discussed in section \ref{sec:spad_device}, the value of $V_\mathrm{bd}$ can potentially be reduced below 30~V by decreasing the radius of the cathode. To maintain a competitive timing resolution, the pitch and depth should also be reduced to values around 7~µm. This roughly corresponds to the dimensions of the SPAD by Ito \textit{et al.} \cite{ito2020}. Therefore, similar PDE could be expected. (ii) Secondly, the DCR of the presented device is higher than the state of the art. The main contribution to the DCR at room temperature appears to be thermal generation. The DCR can likely be improved by optimizing the doping profile, by lowering the field strength near the top interface, and by improving surface passivation. Note that the reported DCR is already low enough for LIDAR applications, for which solar radiation dominates the noise \cite{rablau2019}. (iii) Thirdly, the measured series resistance is higher compared to other devices. Improving this parameter is likely beneficial for power consumption and quenching behavior \cite{cova1996}. The series resistance can be reduced by decreasing the spacing between the device electrodes and by improving the locations of electrical contacts. (iv) Finally, the optical crosstalk of the presented SPAD is high. The electrical crosstalk and spatial optical crosstalk are likely also relatively high \cite{vansieleghem2021iisw}. All forms of crosstalk can be reduced significantly by introducing DTI between devices \cite{ito2020}.

\section{Conclusion}
A BSI silicon SPAD with a charge-focusing electric field was fabricated and characterized. The device operates with a CMOS readout circuit and achieves a PDE of 27\% at 905~nm, low noise, and 240~ps timing resolution. Due to the charge focusing functionality, the detector has the potential to be scaled without significant loss of performance. As a result of these features, the SPAD architecture forms a promising candidate for the integration into large format ToF LIDAR imaging arrays. The potential future developments include the reduction of the breakdown voltage, pitch, and crosstalk, as well as the integration into dense arrays.

\section*{Acknowledgment}
The authors thank Vasyl Motsnyi at imec, Belgium, for help with the quantum efficiency measurements.

\appendix
\section{Crosstalk model}
\label{ap:crosstalk}
Consider a line array with $N_\mathrm{spad}$ identical incrementally-numbered SPADs. The $i$-th SPAD has a primary breakdown event rate of $\nu_{\mathrm{prim},i}$ (as defined in section \ref{sec:crosstalk}). Each primary event in SPAD~$i$ can trigger a secondary event in SPAD~$j$ by direct optical crosstalk, with probability $\mathrm{P}(i\rightarrow j)=\chi_{|j-i|}$. The secondary event may initiate crosstalk by itself, resulting in a crosstalk chain. The probability for a chain is equal to the product of the probabilities of the individual events. For example, $\mathrm{P}(1\rightarrow2\rightarrow4)=\chi_1\chi_2$ represents the chance that a primary event in SPAD~1 triggers a secondary event in SPAD~2 (with probability $\chi_{|2-1|}$), which subsequently triggers a ternary event in SPAD~4 (with probability $\chi_{|4-2|}$). 

Crosstalk in a line array is analyzed under the following conditions: (i) Primary events of all SPADs have a small probability of occurring simultaneously. Therefore, the effect of any primary event can be investigated independently. (ii) When a SPAD enters avalanche breakdown, it can trigger all other SPADs that are not yet in the dead period. (iii) Crosstalk occurs on a faster time scale than the dead time. Therefore, each SPAD can trigger only once as the result of any primary event. (iv) Multiple crosstalk chains can exist at the same time between different SPADs following any single primary event. Not all transitions in these chains are mutually exclusive. Consequently, the probability that multiple chains occur simultaneously (indicated by $\wedge$) equals the product of the probabilities of the unique transitions of those chains. For example, $\mathrm{P}(1\rightarrow2\rightarrow4 \wedge 1\rightarrow2\rightarrow3\rightarrow4)=\chi_1\chi_2\chi_1\chi_1$, wherein the probability for $1\rightarrow2$ is considered only once.

The triggering probability $P_{ik}$ defines the total chance that a primary event in SPAD~$i$ triggers an event in SPAD~$k$. This probability is given by the disjunction
\begin{align*}
\begin{aligned}
P_{ik} &\equiv \mathrm{P}(\text{Any crosstalk chain from SPAD $i$ to SPAD $k$}) \\
&\equiv \mathrm{P}(c_1 \vee c_2 \vee ... \vee c_{M}),
\end{aligned}
\end{align*}
where $\{c_m | m \in \{1,..,M\}\}$ is an exhaustive set of crosstalk chains between the two detectors. To calculate this probability, the disjunction is converted to a sum of conjunctions \cite{jaynes2003}. For two chains $c_1$ and $c_2$, the disjunction can be written as
\begin{align*}
\begin{aligned}
\mathrm{P}(c_1 \vee c_2) =\mathrm{P}(c_1) + \mathrm{P}(c_2) - \mathrm{P}(c_1 \wedge c_2).
\end{aligned}
\end{align*}
Adding an additional chain $c_3$ yields
\begin{align*}
\begin{aligned}
\mathrm{P}(c_1 \vee c_2 \vee c_3) = \mathrm{P}(c_1) + \mathrm{P}(c_2) + \mathrm{P}(c_3) - \mathrm{P}(c_1 \wedge c_2)\\
- \mathrm{P}(c_1 \wedge c_3) - \mathrm{P}(c_2 \wedge c_3) + \mathrm{P}(c_1 \wedge c_2 \wedge c_3).
\end{aligned}
\end{align*}
In general, the probability for the disjunction of $M$ chains is found by alternately adding and subtracting the probabilities of the conjunction of any combination of chains in subsets with increasing sizes. The computational complexity is factorial in $M$. To speed up calculations, only the $M'$ most likely chains are considered, resulting in a slight underestimate of $P_{ik}$.

The total event rate of the $i$-th SPAD is equal to the sum of the primary rate $\nu_{\mathrm{prim},i}$ and the crosstalk contributions from the other SPADs. For independent primary events, it follows that
\begin{equation*}
\nu_{\mathrm{tot},i} = \sum_{k=1}^{N_{\mathrm{spad}}} P_{ik} \times \nu_{\mathrm{prim},k},
\end{equation*}
wherein $P_{ii}=1$ and $P_{ik} = P_{ki} = P_{(N_{\mathrm{spad}}+1-i)(N_{\mathrm{spad}}+1-k)}$. The validity of this equation has been verified with Monte Carlo simulations of the crosstalk process. Using the least-squares fitting procedure, the crosstalk probabilities $\chi_{n}$ can be extracted from the measured event rates in line arrays.


\end{document}